\documentclass[preprint,12pt]{elsarticle}

\usepackage{amssymb,ifpdf,graphicx,color}

\usepackage{ifpdf}

\ifpdf\usepackage{epstopdf}\fi

\journal{Journal of Magnetism and Magnetic Materials}
 
\begin{document}

\begin{frontmatter}

\title{Dilution effects in spin 7/2 systems. The case of the antiferromagnet GdRhIn$_5$}

\author[label1]{R. Lora-Serrano}

\address[label1]{Instituto de F\'isica, Universidade Federal de Uberl\^andia, 38408-100, Uberl\^andia, MG, Brazil}

\author[label2]{D. J. Garcia}

\address[label2]{Consejo Nacional de Investigaciones Cient\'ificas y T\'ecnicas (CONICET) and Centro At\'omico Bariloche, S.C. de Bariloche, R\'io Negro, Argentina}

\author[label3]{D. Betancourth}

\address[label3]{Centro At\'omico Bariloche and Instituto Balseiro (U. N. Cuyo), 8400 Bariloche, R\'io Negro, Argentina}

\author[label1]{R. P. Amaral}

\author[label1]{N. S. Camilo}

\author[label4]{E. Est\'evez-Rams}

\address[label4]{Instituto de Ciencias y Tecnolog\'ia de Materiales, Universidad de la Habana (IMRE), San L\'azaro y L, CP 10400 La Habana, Cuba}

\author[label5]{L. A. Ortellado G. Z.}

\address[label5]{Instituto de Qu\'imica, Universidade Federal de Uberl\^andia, 38408-100, Uberl\^andia, MG, Brazil}

\author[label6]{P. G. Pagliuso}

\address[label6]{Instituto de F\'isica ``Gleb Wataghin'', UNICAMP, 13083-970, Campinas-S\~ao Paulo, Brazil}

\begin{abstract}

We report the structural and magnetic characterization of La-substituted Gd$_{1-x}$La$_x$RhIn$_5$ ($x\leq$ 0.50) antiferromagnetic (AFM) compounds. The magnetic responses of pure GdRhIn$_5$ are well described by a $S=7/2$ Heisenberg model. When Gd$^{3+}$ ions are substituted  by La$^{3+}$, the maximum of the susceptibility and the inflection point of the magnetic specific heat are systematically shifted to lower temperatures accompanied by a broadening of the transition. The data is qualitatively explained by a phenomenological model which incorporates a distribution of magnetic regions with different transition temperatures ($T_N$). The universal behaviour of the low temperature specific heat is found for La (vacancies) concentrations below $x=0.40$ which is consistent with spin wave excitations. For $x=0.5$ this universal behaviour is lost. The sharp second order transition of  GdRhIn$_5$ is destroyed, as seen in the specific heat data, contrary to what is expected for a Heisenberg model. The results are discussed in the context of the magnetic behavior observed for the La-substituted (Ce,Tb,Nd)RhIn$_5$ compounds.

\end{abstract}

\begin{keyword}

Antiferromagnetism \sep Magnetic dilution \sep Substitutional disorder \sep Heisenberg model



\end{keyword}

\end{frontmatter}

\section{Introduction}
\label{introd}

The intermetallic compound GdRhIn$_{5}$ belongs to the family $R_{m}M_{n}$In$_{3m+2n}$ (\textit{R} = Ce-Tb; \textit{M} = Rh, Ir or Co; \textit{m} = 1, 2; \textit{n} = 0, 1). When $R =$ Ce, heavy fermion behaviour, Kondo effect, quantum criticality, anomalous metallic behaviour and magnetic order combine with unconventional superconductivity (USC) in the phase diagram (see Ref. \cite{Sarrao1} and references therein). 
The crystal structures along the series depend on the number of \textit{m} layers of cubic $R$In$_3$ units stacked sequentially along the \textit{c} axis with intervening $M$In$_2$ layers. Single layer members crystallize in the tetragonal HoCoGa$_5$-type structure.

To understand the evolution of the magnetic properties, an instructive exercise is to evaluate the strength of the magnetic interaction to dilution by substituting the magnetic rare earth atoms with non-magnetic ones. For instance, in the Ce$_{1-x}$La$_{x}$RhIn$_{5}$ and Ce$_{1-x}$La$_{x}$CoIn$_{5}$ the long range Ruderman-Kittel-Kasuya-Yoshida (RKKY) magnetic interaction is affected in different ways\cite{pagliuso7,VictorCorrea,Light,Alver,christianson2,wei3,Tanatar,Petrovic3,Satoro,Leticie}.
The magnetic order along the series Ce$_{1-x}$La$_{x}$RhIn$_{5}$ is suppressed at a critical concentration compatible with a two dimensional magnetic order, due to the anisotropic couplings\cite{pagliuso7,Light,wei4,Takeuchi2006}. The evolution of the magnetic properties within the non-Kondo La-substituted series Tb$_{1-x}$La$_{x}$RhIn$_{5}$ and Nd$_{1-x}$La$_{x}$RhIn$_{5}$ is different.
Because of the crystal field effects, these compounds behave like Ising antiferromagnets with sharp second order transitions as seen on specific heat measurements\cite{raimundo5,raimundo7}.
For the Tb series, $T_{N}$ extrapolates to zero at about 70\% of La content which is compatible with a three-dimensional order.
In these cases, the magnetic susceptibility and specific heat data can be modelled by considering the evolution of the magnetic exchange and crystal field parameters with doping. 

The substitution of the magnetic \textit{R} ion by La introduces vacancies in the magnetic lattice. These vacancies can induce weak or, the so-called, quenched disorder \cite{vojta3}.
The effect of this disorder depends on the lattice symmetry and the dimensionality of the interactions.
When the vacancies concentration is larger than the percolation limit the magnetic order is completely suppressed.
This percolation limit has been determined for various models, in particular for classical spins on cubic \cite{MagnPhTrans,Kato,Vojta1} and square \cite{Yonezawa} lattices.
Further, the character of the paramagnetic (PM) to antiferromagnetic AFM transition may change.
For Ising or classical Heisenberg models on a cubic lattice, few impurities do not change the sharp second order transition\cite{vojta3}. Instead, renormalized parameters or critical exponents can be defined. This corresponds to the behaviour seen on the Tb and Nd series\cite{raimundo5,raimundo7}.

In this work, we study the Gd$_{1-x}$La$_x$RhIn$_5$ series with different Lanthanum concentration. GdRhIn$_5$ undergoes an AFM transition below $T_{N}\approx$ 40 K to a commensurate magnetic structure with propagation vector $(0,\frac{1}{2},\frac{1}{2}$) and the moments oriented in the tetragonal \textit{ab}-plane\cite{pagliuso3,granado2}. 
Along the La-substituted Nd-, Tb- and Ce-based compounds, first order crystal field effects, anisotropic couplings as well as magnetic dilution effects take account of the $T_N$ decrease with $x$\cite{pagliuso7,pagliuso5,christianson1,raimundo5,raimundo7}. Gd$^{3+}$ is a magnetic \textit{S}-ion which cannot be affected by first order crystalline electric field (CEF) effects (because of its null angular momentum), nor by anisotropic couplings. Instead, only dilution effects should be present. However, there are other sources of magnetic anisotropy, as the magnetic dipolar interaction\cite{JensenAndMackintosh,granado2}.  
Recently, the magnetic couplings in the compounds Gd$M$In$_5$ ($M$= Co,Rh,Ir) were determined through first principles calculations to be approximately homogeneous\cite{Facio2015}. 

In order to describe the dilution effects and their relevance to the physics of the $R$RhIn$_5$ family (R=Ce,Tb,Nd,Gd), we show that GdRhIn$_5$ is well described by a $J$= 7/2 Heisenberg model. When Gd is substituted with Lanthanum, the thermodynamic data can be compared to a phenomenological model to account for magnetic disorder. The latter is responsible for the broadening of the specific heat data around $T_N$, as well as for the abrupt decrease of $T_N$ with $x$. Disorder driven short-range order has been proposed for Ce$_{0.5}$La$_{0.5}$RhIn$_5$\cite{Light,pagliuso7}.

The paper is organized as follows. In Section \ref{experim}, we present the results from the analysis of the powder diffraction data and magnetic characterization as a function of temperature (T) of the title compounds. 
In Section \ref{teoria}, the magnetism of GdRhIn$_5$ is theoretically discussed. A model to account for the magnetic disorder is introduced for the doped cases. The specific heat data for $x\leq0.4$ can be rescaled into a universal curve for low T. In Section \ref{discusion}, the experimental results are discussed in the framework of the classical Heisenberg model by considering the large Gd$^{3+}$ spin value ($S$=7/2). Our results are compared to those obtained within other La-substituted non-S members from the $R$RhIn$_{5}$ series, where CEF effects are relevant. Finally, we present our conclusions.

\section{Experimental}
\label{experim}

Single crystals of the Gd$_{1-x}$La$_{x}$RhIn$_{5}$ (nominal concentration \textit{x} = 0.00, 0.15, 0.30, 0.40 and 0.50) compounds were synthesized by the metallic flux technique \cite{Fisk1}. High purity chemical elements were prepared in the (1-x):(x):1:20 proportion for Gd:La:Rh:In, placed in an alumina crucible and sealed in vacuum in a quartz tube before being treated in a conventional furnace.  The tetragonal HoCoGa$_5$-type structure (space group $P4/mmm$) were confirmed by ambient temperature X-ray powder diffraction (XPD) data taken in a Shimadzu XRD-6000 diffractometer working in the Bragg-Brentano geometry, graphite monochromator and Cu K$_{\alpha}$ radiation. For the magnetic characterization, \textit{dc} susceptibility measurements were performed in a commercial SQUID magnetometer at H = 1 kOe in the temperature range between 2 and 300 K. Specific heat measurements were performed in a commercial small-mass calorimeter that employs a quasi-adiabatic thermal relaxation technique. C(T) data were taken between 1.8~$\leq$~T~$\leq$~60~K.

\begin{figure}
\centering
\includegraphics[width=0.75\textwidth]{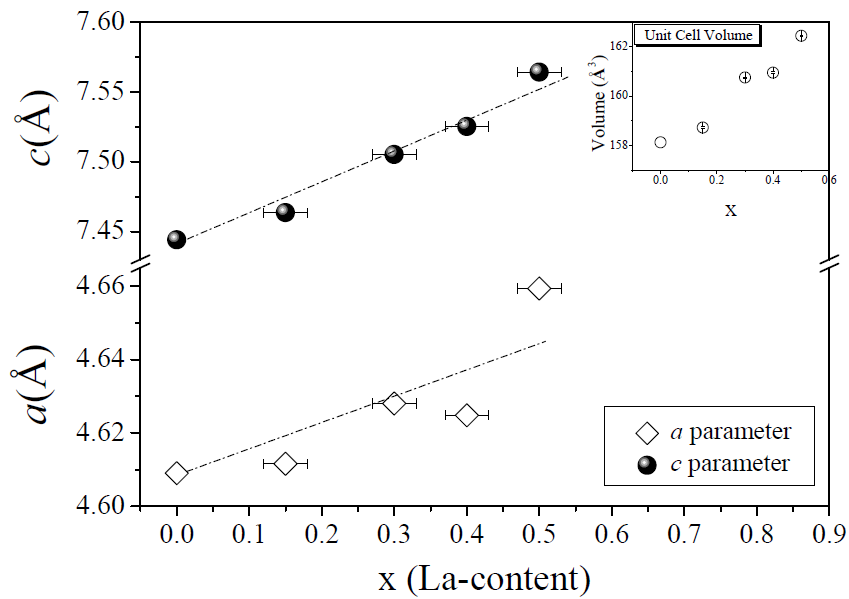}
\caption{\label{Fig1DRX}(a) Unit cell parameters evolution as a function of $x$. Error bars represent the concentration uncertainty ($\Delta x=$0.03), as extracted from susceptibility data (below). Dashed dot lines are guides to the eyes. The evolution of the unit cell volume with $x$ is depicted in the inset.}
\end{figure}

\subsection{Experimental Results}
\label{disc}

Fig.~\ref{Fig1DRX} shows the evolution of the tetragonal \textit{a} and \textit{c} cell parameters as the La-content is increased. The unit cell volume evolution can be seen in the inset. Both behaviours follow the Vegard's law for solid solutions. \textit{a}, \textit{c} and unit cell volumes were determined from least-squares fittings of the Bragg peak positions \cite{Holland}. Statistical error bars for the cell parameters (vertical) are smaller than the symbols used and cannot be observed. Horizontal error bars were estimated from linear fits to the inverse of the magnetic susceptibility data (see below). 
For the single crystal orientation, Laue diffraction data was taken for all studied sample. From this data is possible to confirm that the sample surface is perpendicular to the [001] direction (plate-shape morphology).

\begin{figure}[ht!]
\centering
        \includegraphics[width=0.68\textwidth]{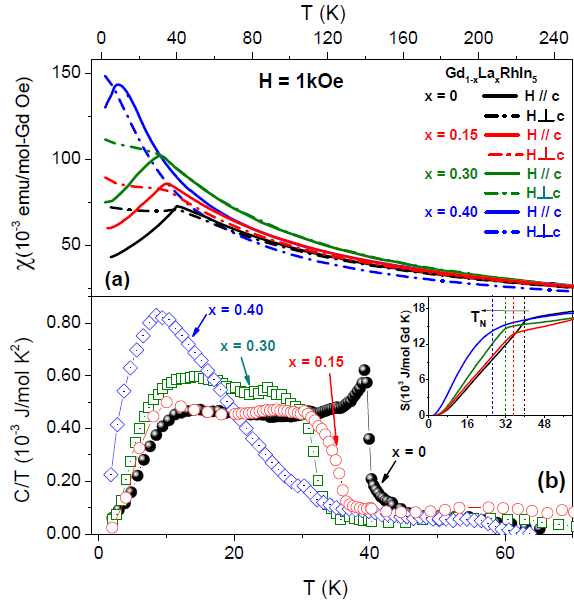}
    \caption{\label{Fig2ChiCp}(Color online) (a) Magnetic susceptibility $\chi_{\parallel}$ (continuous curve) and $\chi_{\perp}$ (dash-dotted curve) as a function of temperature and La-content. (b) Magnetic contribution to the specific heat as a function of temperature, C/T(T). The inflection points from the C/T anomaly (high T side) were defined as $T_N$. Data for each La concentration is indicated by arrows. The inset shows the magnetic entropy (see text).}
\end{figure}

Fig.~\ref{Fig2ChiCp}(a) shows the temperature dependence of the \textit{dc} magnetic susceptibility with the magnetic field perpendicular, $H\perp c$, and parallel, $H\parallel c$, to the tetragonal \textit{c}-axis. For the perpendicular measurements, the field was first applied parallel to the two in-plane directions (\textit{a} and \textit{b}) -- not shown. No differences were observed between both datasets. Hence, the measurement along one of these directions was assumed as the $\chi_{\perp}(T)$ data for each concentration. In order to extract the effective magnetic moment per Gd$^{3+}$ ion ($\mu_{eff}$) and estimates the La concentration, the average polycrystalline susceptibility $\chi_{poly}$ = ($\chi_{//}$+2$\chi_{\perp}$)/3 was determined (not shown). From the linear fit to high temperature data of the reciprocal $\chi_{poly}^{-1}$ vs. \textit{T} data, $\mu_{eff}$ was calculated within $\pm 0.03$ of uncertainty (horizontal error bars in fig.~\ref{Fig1DRX}). 

Fig.~\ref{Fig2ChiCp}(b) shows the magnetic contribution to the specific heat data, C/T, for zero applied field ($x\leq0.40$). The inflection point in the C/T curves coincides well with the maximum for $\chi$($T$) data for $x=0, 0.15$ and $0.30$. Therefore, the T values extracted from C/T data have been taken as N\'eel temperatures $T_N$ for $x<0.40$. For \textit{x} = 0.50 (not shown) no anomalies were observed within the studied T range. All curves  in Fig.~\ref{Fig2ChiCp}(b) were corrected for the phonon contribution using the non-magnetic C/T data of LaRhIn$_5$. The magnetic entropy, calculated as $\int^{T_{max}}_{T_{min}} C/T dT$, was also extracted (see inset; $T_{min,max}$ refers to the minimum and maximum measured temperatures). Around $T_{max}$ all the expected $3Rln2$ entropy per mol of Gd is recovered.

\section{Modeling weak disorder on a $S=7/2$ spin lattice}
\label{teoria}

In this section, we first show that both the $\chi_{poly}$ and C(T) data of GdRhIn$_5$ are well reproduced by the Heisenberg model in a cubic lattice. We use this result to introduce a simple, phenomenological, model for quenched disorder that accounts for the experimental results for small La concentrations. In particular, substitutional effects are described by a spin model with a single energy parameter.

\subsection{Susceptibility and specific heat of the $S=7/2$ 3D Heisenberg model}
\label{QMC}
\begin{figure}[htbp]
\centering
        \includegraphics[width=0.8\textwidth]{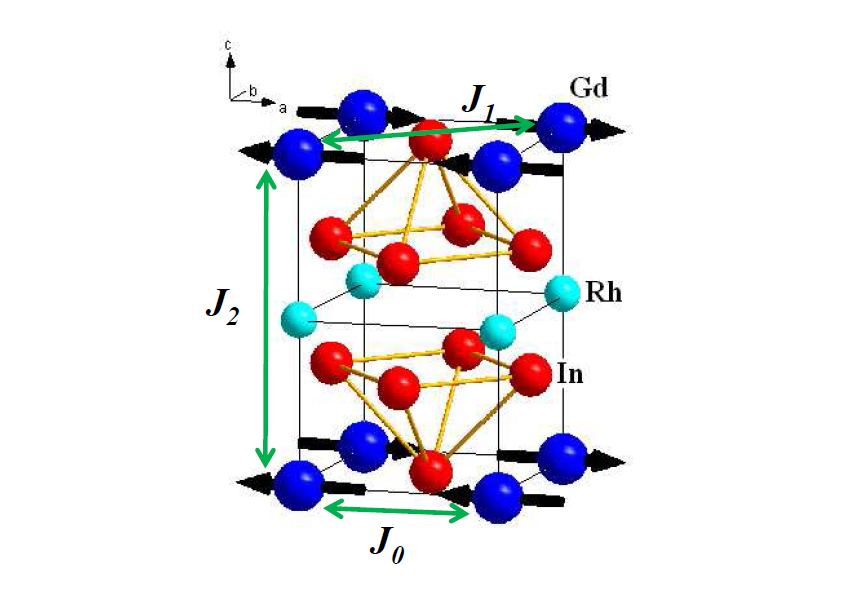}
    \caption{\label{Fig3UnitCell}(Color online) The \textit{C}-AFM magnetic structure of GdRhIn$_5$. $J_0$, $J_1$, and $J_2$  magnetic exchange between Gd atoms are indicated by double sided arrows. Solid arrows represent the relative orientation of the Gd magnetic moments according to the wave vector $\vec{k}= (0\frac{1}{2}\frac{1}{2})$ and the orientation as determined in ref. \cite{granado2}.}
\end{figure}

Magnetic correlations between rare earth moments within the \textit{R}RhIn$_5$ series have been discussed in refs. \cite{pagliuso3,Takeuchi2006,granado2}. It is known that they couple at first order through three different Heisenberg ($J_i$) terms (Fig.~\ref{Fig3UnitCell}). There are two couplings between nearest neighbours (NN) in the $ab$ plane: a term $J_0$ along the [100] direction and a $J_1$ term between next-nearest neighbours (NNN) in the plane diagonal [110]. A third contribution, $J_2$, takes place between NN along the [001] direction\cite{Takeuchi2006}. The existence of both $J_0$ and $J_1$ couplings induces magnetic frustration. When $J_1>0.6 J_0$, this frustration is partially solved with ferromagnetic (FM) chains along one of the crystallographic directions, and AFM correlations along the two others \cite{granado2,Bacci,Sushkov}. Single arrows in Fig.~\ref{Fig3UnitCell} indicate the relative orientation of the Gd magnetic moments according to the wave vector $\vec{k}= (0\frac{1}{2}\frac{1}{2})$\cite{granado2}. For the Gd$M$In$_5$ compounds ($M$=Rh,Co,Ir), first principle calculations have estimated the values of these couplings. It allows to quantitatively explain several thermodynamics measurements \cite{Facio2015,DBetancourth1}.

\begin{figure}[htbp]
\centering
        \includegraphics[width=0.65\textwidth]{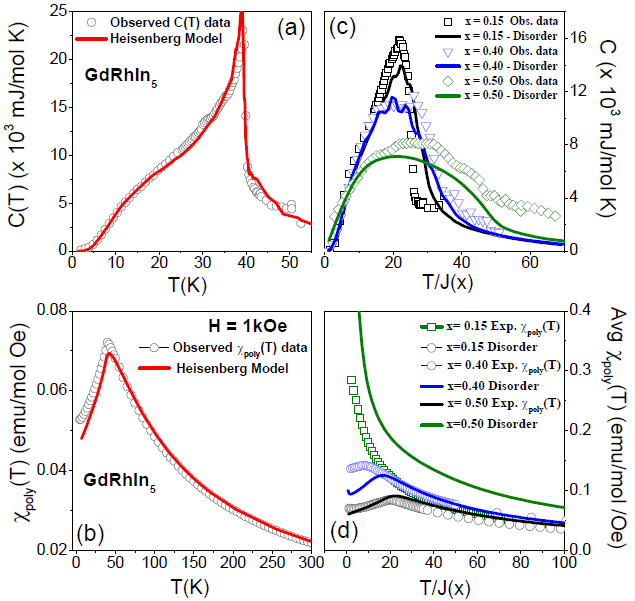}
    \caption{\label{Fig4ChiCpCalc} (Color online) C(T) vs. T (a) and the $\chi_{poly}$ (b) data for the GdRhIn$_5$ compound compared to the 3D Heisenberg model for spin $S=7/2$. Panels (c) and (d) show representative observed data together with the discrete (\textit{x} = 0.15, 0.40) and continuous (\textit{x} = 0.50) averaged disorder simulations (see section~\ref{SubsDis}). For the simulated data we used continuous curves to represent \textit{x} = 0.15 (black), 0.40 (blue) and 0.50 (green) concentrations. The normalized T/J(x) scale is used to allow the comparison between the C(T) low temperature data. J(x) = 1.7, 0.8 and 0.7 K for $x$ = 0.15, 0.40 and 0.50, respectively.}
\end{figure}

The specific heat and the $\chi_{poly}$ data of GdRhIn$_5$ are reproduced by a Heisenberg model considering only a single coupling  $J$ between first neighbours (Figs.~\ref{Fig4ChiCpCalc}(a) and (b)). This coupling represents the average of the $J_0$, $J_1$ and $J_2$ terms. For GdRhIn$_5$, $J=1.83$ K, as determined by first principle calculations \cite{Facio2015}. To fit the experimental data, here we used a slightly smaller $J=1.7$ K, similar to what has been done in ref. \cite{Facio2015}. 

The Heisenberg model in a cubic lattice is given by the Hamiltonian

\begin{equation}
H= J\sum_{\langle i,j \rangle} S_i \cdot S_j, 
\label{eq:HHeisenberg3D}  
\end{equation}
where $S_{i,j}$ are quantum spins with $\vert S\vert=7/2$.  This model has a check-board like ground state (G-AFM) contrary to the actual C-AFM order of GdRhIn$_5$. However, the specific heat and polycrystalline susceptibility are well reproduced.

To solve the model we have used a quantum Monte Carlo (QMC) algorithm from the ALPS library \cite{ALPS} for $T>10$ K. In particular, we use the ``loop'' algorithm with up to $10^6$ thermalization and sweep steps and compare with  numerical results extrapolated to the thermodynamic limit. For disorder averages we use results obtained in a $12^3$ lattice (see below).
 QMC error bars (not shown in Fig \ref{Fig4ChiCpCalc}) are roughly of the same order of the symbol sizes used for the observed data. 
At low temperatures, AFM spin waves for the specific heat were used:

\begin{equation}
C (T\rightarrow 0) = R\frac{32\times\pi^5}{15 12^{3/2}} (\frac{T}{S J})^3,
\label{eqSpinWaves}
\end{equation}
where \textit{R} is the gas constant. A gyromagnetic factor $g$ value of 2 was assumed for susceptibility.
For the full derivation of the specific heat for all T ranges see ref.~\cite{Facio2015}.

\subsection{Simulating the substitutional disorder} 
\label{SubsDis}

La substitution at the \textit{R} sites introduces vacancies in the magnetic lattice which are randomly distributed. This kind of disorder is called weak or quenched disorder, as it does not evolve with time nor induce frustration \cite{vojta3}. These vacancies modify the number of magnetic neighbours, which in turn induces spatial variations on the coupling strength.
In some situations, the magnetic transition remains sharp in the presence of these defects, as in the cases of the Ising and Heisenberg model in a cubic lattice. But when ``rare regions'' are present, the sharp phase transition can be completely destroyed \cite{vojta3}. These are regions where vacancies are absent or, more generically, where an approximately uniform field acts on every spin.
As shown above, doping GdRhIn$_5$ with La changes the sharp transition to a smooth one (Figs.~\ref{Fig2ChiCp} and~\ref{Fig4ChiCpCalc}). 
This experimental fact indicates that ordered regions are an important ingredient to model substitutional disorder.
For each of these ordered regions a critical temperature can be defined. The upper limit for the vacancy concentration that destroy the magnetic order is given by the 3D magnetic percolation limit. Above this concentration ($x_{c(3D)}=0.68$ for classical spins on a cubic lattice~\cite{MagnPhTrans,Kato,Vojta1}) most magnetic ions are uncorrelated. Below this limit the magnetic order is usually well defined at low temperatures.

A complete theoretical study of the effect of vacancies on the magnetic model would imply numerical simulations of lattices with randomly removed magnetic sites in the compound full model, probably a Heisenberg model with at least NNN hopping. Such kind of computational approximation (see for instance ref. \cite{Kato}) is beyond the scope of this work. Instead, we introduce a simple, phenomenological model for the disorder originated by vacancies. 
We assume that 
\begin{itemize}

\item The disordered lattice has regions where the number of neighbours $y$ is approximately constant. 
This number varies between $0$ and the number $z$ of magnetic neighbours in the non-substituted compound GdRhIn$_5$.
Thermodynamic quantities follow from a weighted average of the properties for those regions. 
This hypothesis should fail close to the percolation limit as a fractal dimension gives a more accurate description of the disordered lattice \cite{Vojta1}.

\item The Hamiltonian for a region with $y$ neighbours is the same as for a 3D Heisenberg model on a simple cubic lattice. Only the coupling strength is rescaled according to the ratio $\frac{y}{z}$. 
This approximation mimics the mean field approach, where only the effective field $y J$ acting on a site determines all properties (see bellow), and is partially justified by the successfully application of the 3D Heisenberg model on a simple cubic lattice (previous section). It is worth noting that the $z=6$ describes the cubic case, but the real experimental situation of tetragonal symmetry should take into account 10 couplings (Fig. 3): the 6 NN (with $J_0$,$J_2$ coupling) and 4 NNN (with $J_1$).
However, when $y$ is small ($y\lesssim 2$), it can be inaccurate as the geometry for that neighbourhood is very different.
%It can be very poor  as the geometry for that number of neighbours is very different.

\end{itemize}

\begin{figure}[t!]
\centering
        \includegraphics[width=0.6\textwidth]{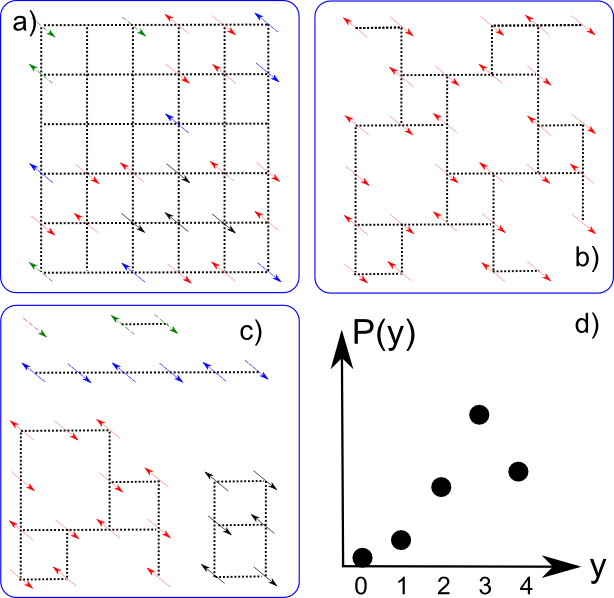}
    \caption{\label{MagnDisorder}(Color online) Analysis of the magnetic disorder in a two dimensional case. (a) Example of a network with one magnetic vacancy every four sites; 1-, 2-, 3- or 4-neighbour sites are represented by green, blue, red and black arrows, respectively. (b) Homogeneous situation where every site has three neighbours. (c) Collection of regular arrangements with 0, 1, 2, 3 or 4 neighbours extracted from (a). (d) Probability of each spin configuration vs. number of neighbours.}
\end{figure}

In order to illustrate our approach, in Fig.~\ref{MagnDisorder} we present a two dimensional example of this analysis of disorder. We consider spins on a square lattice with one vacancy every four sites. In this network of spins, there will be isolated (with no magnetic neighbours), 1-, 2-, 3- or 4-neighbour sites.
The last four cases are highlighted, respectively, with green, blue, red and black arrows in Fig.~\ref{MagnDisorder} (a). Fig.~\ref{MagnDisorder} (b) considers the homogeneous situation where every site has the average value of three neighbours. A slightly more complex approach to the real disorder can be considered by decomposing the spin network of panel (a) as a collection of regular arrangements with $y=0, 1, 2, 3$ or 4 neighbours (Fig.~\ref{MagnDisorder} (c)).
Each situation has a given probability (panel d) which, for simplicity, we approximate to a binomial distribution. In this case, the thermodynamic properties are {\it weighted} averages of rescaled results from the dense spin network. Thermodynamic results for $y=0,1,2,3,4$ are taken from those of the original dense lattice with a coupling $y/4J$.
%This last approach, for low La substitution, is very close to the experimental data (Figs.~\ref{Fig4ChiCpCalc}(c) and (d)above).

To gain insight into the effect of disorder in the 3D case of Gd$_{1-x}$La$_x$RhIn$_5$, let us begin with a mean field approximation.
We consider an antiferromagnet with only NN interaction.
The effective field on each site is given by $B_{eff}\propto z J \langle S \rangle$.
$z$ is the number of NN, $J$ the NN coupling and $\vert\langle S \rangle\vert$ is the mean value of the spin on any site. If a fraction $x$ of sites become vacant (by substituting magnetic ions by non-magnetic ones, for instance), the average number of neighbours decreases from $z$ to $z (1-x)$ and the effective field acting on a site decreases correspondingly. The $B_{eff}$ strength determines the paramagnetic to antiferromagnetic phase transition (N\'eel) temperature $T_N=z (1-x) \frac{S(S+1)}{3}J$ \cite{BlundellMagnetism}.
For $x<1$, this transition is always second order with a sharp discontinuity in the specific heat.
Certainly, this is not the experimental situation observed in the series Gd$_{1-x}$La$_x$RhIn$_5$ for $x>0$, which shows that a better approach must be implemented.

We now consider the existence of regions where order can be defined. 
Inside each region, any site has exactly $y$ neighbours. This region appears with a relative weight or probability $P(y)$.
We approximate $P(y)$ as the probability of finding $y$ magnetic neighbours in a site of the full sample. 
Each site could have up to $z$ magnetic neighbours, thus $0\leq y\leq z$. 
The probability of a given site to have $y$ spin neighbours in a given configuration is the product of the probability of having exactly $y$ spin neighbours and the probability of having $z-y$ vacancies, $(1-x)^y x^{z-y}$. 
The number of configurations is given by the combinatorial of $y$ on $z$, ${y \choose z}$. 
Therefore, the probability of having a region with $y$ spin neighbours is given by the binomial probability $P(y)= {y \choose z} (1-x)^y x^{z-y}$. 

The sample is represented by the original lattice  with $z$ neighbours, but the coupling is rescaled $J_{eff}(y) = y/z J$.
This rescaling assures that the effective field in the region is proportional to $y J$ and relates the thermodynamic properties for each region (with $y$ neighbours) to those of the original model with $z$ neighbours and coupling $J$.
This approximation is clearly wrong for small values of $y$.
For example, $y=0$ or 1 would mean an isolated atom or a two atoms molecule, respectively, while $y=2$ implies a linear chain, which cannot give a finite ordering temperature.
%drive to a 3D thermodynamic response.
 We assume that for $x\leq 0.50$ such small number of neighbours ($y\leq 2$) is uncommon and have low impact in the specific heat.
For the magnetic susceptibility, we consider that these regions have a free ion paramagnetic contribution.

The computation of the disorder-averaged specific heat and susceptibility data follows below. The subindex ``$y$'' (``$z$'') will be used for the sample (original) lattice property. For comparison with the experimental results we use $z=10$, which corresponds to the number of relevant magnetic neighbours in GdRhIn$_5$ (see fig.\ref{Fig3UnitCell}).

\subsubsection{Specific heat.}
\label{SpecHeat}

The specific heat for each region with $y$ neighbours and effective coupling $J_{eff}(y)$ can be related to the original lattice by
\begin{eqnarray}
C_y(T)&=& \frac{d E_y(T)}{d (T/J_{eff}\times J_{eff})} \\ \nonumber
&=& \frac{d E_y(T)/J_{eff}}{d (T/J_{eff})}.\\ \nonumber
\end{eqnarray}

Because there is a single energy scale $E_y(T=\alpha J_{eff})/J_{eff} = E_z(T=\alpha J)/J$, so 

\begin{eqnarray}
C_y(T)&=& \frac{d E_z(T/J_{eff}\times J)/J}{d (T/J_{eff})}\\ \nonumber
&=& \frac{d E_z(T/J_{eff}\times J)}{d (T/J_{eff}\times J)}\\ \nonumber
&=&C_z(T/J_{eff}\times J), \nonumber
\end{eqnarray}
meaning that the specific heat in a region equals the specific heat in the original model at temperature $T/J_{eff} \times J$.

For a collection of samples with couplings $J_{eff}(y)$ and relative weight $P(y)$ the averaged specific heat is given by
\begin{equation}
C(T)=\sum_{y=1}^z C_y(T) P(y)=\sum_{y=1}^z C_z(\frac{T}{J_{eff}(y)} \times J) P(y),
\end{equation}
where $C_z$ is the specific heat for the ordered ($x=0$) Hamiltonian. Notice that $C_z$ could be obtained from the experimental data of the non substituted GdRhIn$_5$ compound. Then, we start from the Heisenberg model specific heat of GdRhIn$_5$ as it correctly describes the experimental data (Fig.~\ref{Fig4ChiCpCalc} (b)).

The results from these averages are shown in Fig.~\ref{Fig4ChiCpCalc}(c) with continuous curves. Black, blue and green colours were used to identify the curves for the $x = 0.15, 0.40$ and $0.50$, respectively.
The temperature has been scaled by the average exchange term $J(x)=(1-x)\times J$. 
A good agreement between the modelled and experimental data can be seen. When \textit{x} increases, the specific heat gets smoother around the transition. In the simulation, the original $x=0$ sharp transition is split between $z$ smaller transitions. This causes the splines observed around the transition for the $x=0.15, 0.40$ continuous curves. 

For $x=0.5$, we computed the specific heat by considering a flat distribution of $y$, $P(y)=1/z$, for continuous values of $0<y\leq z$. This distribution has shown to be more adequate to describe the specific heat and susceptibility data (Fig.~\ref{Fig4ChiCpCalc} (c)). 

For all concentrations and large enough T/J(x) ($\geq 40$) the universal behaviour is recovered. This is also the case for low $T$ and concentrations $x\leq 0.40$. It is worth to notice that despite we only show data for $x=0.15$ and $0.40$, the low T data for $x=0.00$ and $0.30$ also collapses for T/J(x)$ \leq $10.
This universal behaviour corresponds to spin waves ($C\propto T^3$) at low T. 
For $x=0.5$ the spin-wave-like behaviour is lost as a consequence of the large amount of disorder.

\subsubsection{Susceptibility}
Similar to the specific heat calculation, the susceptibility $\chi_y$ for the region with $y$ neighbours can be related to the original lattice susceptibility $\chi_z$ by
\begin{eqnarray}
\chi_{y}(T)&=& \frac{d^2 E_{y}(B,T)}{d (B/J_{eff}\times J_{eff})^2} \\ \nonumber
           &=& \frac{J}{J_{eff}}\chi_z(T/J_{eff}\times J) \\ \nonumber
\end{eqnarray}

The average susceptibility of a collection of samples with $J_{eff}(y)$ and relative weight $P(y)$ is
\begin{eqnarray}
\chi(T)=\sum_{y=0}^z \chi_y(T) P(y)=\sum_{y=0}^2  P(y) (g\mu_B)^2 \frac{S(S+1)}{3 k_B T} + \\ \nonumber
\hspace{40mm} +\sum_{y=3}^z  \frac{J}{J_{eff}(y)}\chi_z(T/J_{eff}(y)\times J) P(y) \\ \nonumber
\end{eqnarray}
where for $y\leq 2$ we consider a paramagnetic susceptibility.

These disorder-averaged susceptibilities are shown in Fig.~\ref{Fig4ChiCpCalc} (d). The temperature has been scaled by the average exchange term $J(x)$. A good agreement between experiment and theory can be observed for the smallest doping $x=0.15$, while for the $x=0.40$ the model agrees well in the paramagnetic region and down to near $T_N$. For $x=0.50$, only a qualitative agreement is seen: the continuous model reproduces the lack of the maximum and the divergence when $T\rightarrow 0$.

\section{Discussion}
\label{discusion}

We have shown that magnetic specific heat and susceptibility data in GdRhIn$_5$ are well reproduced by the Heisenberg model in a cubic lattice. As the La-content increases, the maximum on $\chi(T)$ shifts to lower temperatures in agreement with the weakening of magnetic correlations between Gd$^{3+}$ ions. There is also an increased broadening of the C/T data around the transition temperature.
Broadening and shifting of the anomaly around $T_{N}$ in the specific heat data was also observed in the Ce$_{1-x}$La$_{x}$RhIn$_5$ series\cite{pagliuso7,Light}. For the series Tb$_{1-x}$La$_{x}$RhIn$_5$ and Nd$_{1-x}$La$_{x}$RhIn$_5$, an Ising-like magnetic behaviour was observed even close to the percolation limits\cite{raimundo5,raimundo7}. In an Ising-like system, it is expected that weak disorder renormalizes the exchange parameters and the transition should remain sharp~\cite{vojta3}, as seen for the Tb and Nd based series.
Given that for GdRhIn$_5$ the magnetic properties are well described by the Heisenberg model and their large spin, one should expect that the substitutional disorder affects this system like in the classical Heisenberg model. For the latter, disorder affects the magnetic properties in the same way as in the Ising model, renormalizing parameters and having a sharp magnetic transitions\cite{vojta3}. As this is contrary to our experimental results on doped GdRhIn$_5$, in order to explain the data along the series we consider the presence of a distribution of regions with well defined order.

In this work, we have modelled the weak disorder as a distribution of regions, each with a uniform effective coupling between spins, which causes a distribution of $T_N$. The experimental data for the series Gd$_{1-x}$La$_{x}$RhIn$_5$ ($0.15\leq x\leq0.40$) shows an apparent contradiction with the Heisenberg model. Our model for disorder assumes a larger number of neighbours $z$ than in the original Heisenberg Hamiltonian, which means that the couplings are of longer range. In particular, next-nearest in-plane neighbours have to be taken into account to explain the C-AFM order.
For $x=0.50$, the substitutional disorder affects even the basic mechanisms destroying the low energy spin waves excitations (Fig.~\ref{Fig4ChiCpCalc} (c)). The flat distribution used to account for that case points to  relevant long range interactions, quite probably related to RKKY couplings.

The experimental magnetic specific heat can be well reproduced for all concentrations.
This good agreement reflects the fact that specific heat measurements sense energy variations.
The low T universal behaviour is clearly seen for all concentrations $x\leq0.4$ pointing to the existence of well defined spin wave excitations.
This good agreement is less obvious for the susceptibility data, particularly for $x\geq0.4$. From the theoretical side, the model considers a simple paramagnetic behaviour for regions with small number of neighbours. As these regions are more common for higher La concentration, their contributions are more complex, especially around the transition. This is clear for the $\chi_{poly}$ data of $x=0.5$ where a low T paramagnetic divergence is seen. These divergent contributions certainly affect the existence of a maximum in the susceptibility data at finite temperatures (down to the lowest T available). Divergent susceptibilities when $T\rightarrow 0$ can be obtained also in two dimensional spin arrangements. This change in the interactions dimensionality is the dominant effect in Ce$_{1-x}$La$_x$RhIn$_5$ where inter-plane couplings are inferred to be smaller than in-plane couplings\cite{pagliuso7,Light,wei4}. However, for GdRhIn$_5$, it was shown that all couplings have similar magnitude\cite{Facio2015}. The presence of the spin disorder, as discussed above, may indeed change the dimensionality of interactions with La content, thus affecting the $T_N$ behaviour with $x$.

\begin{figure}
\centering
        \includegraphics[width=0.75\textwidth]{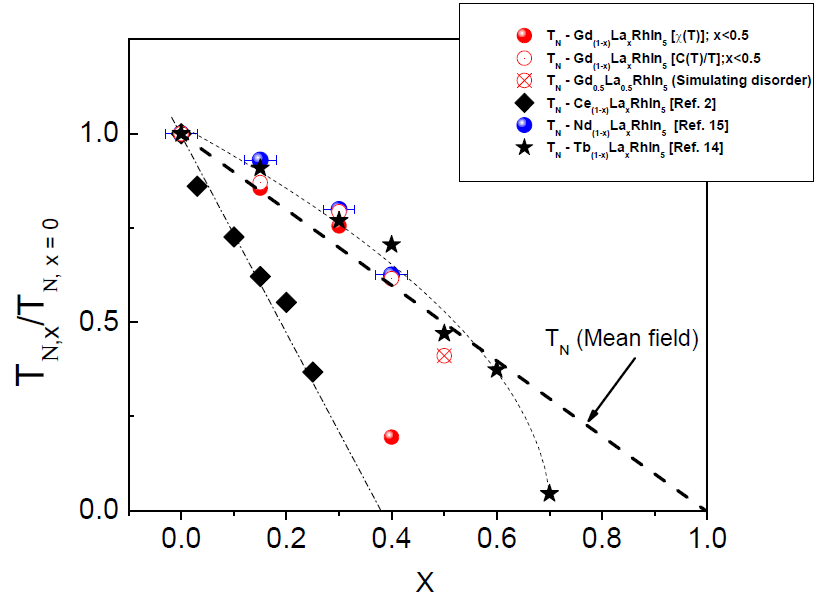}
    \caption{\label{Fig5TNevolution}(Color online) Normalized N\'eel temperature (T$_N$) evolution for the \textit{RE}RhIn$_5$ series (\textit{RE} = Ce, Tb and Gd) as a function of La-substitution. The thick dashed line is the mean field $T_N$ prediction.}
\end{figure}

In Fig.~\ref{Fig5TNevolution} we analysed the evolution of $T_N$ for the Gd$_{1-x}$La$_x$RhIn$_5$ series, normalized by the $T_N$ of the GdRhIn$_5$ compound, $T_{N,x}/T_{N,x=0}$. 
For the Gd-based series, the data were extracted from the results of Fig.~\ref{Fig2ChiCp} for all the studied samples. It can be observed that for $x\leq$ 0.30, $T_N$ can be defined with the same value both from the maximum of the susceptibility or the inflection point of the $C(T)/T$ data. Indeed, up to this concentration $T_N$ varies smoothly following the expected mean field tendency ($T_N \propto 1-x$.)
For larger concentrations both criteria differ, thus $T_N$ defined from susceptibility data drops abruptly for $x = 0.40$ while remains close to the mean field expected value (dashed line) when defined by C/T. For $x=0.50$, the cross-filled circle symbol indicates the normalized $T_N$ as extracted from the disorder simulation. We show for comparison the reported $T_N$ values from the series Ce$_{1-x}$La$_x$RhIn$_5$ (filled diamonds) \cite{pagliuso7}, Tb$_{1-x}$La$_x$RhIn$_5$ (stars) \cite{raimundo5} and Nd$_{1-x}$La$_x$RhIn$_5$ (filled circles with error bars)\cite{raimundo7}. The normalized $T_N$ for Gd$_{1-x}$La$_x$RhIn$_5$, as extracted from specific heat data, closely follows the behaviour of the series Tb$_{1-x}$La$_x$RhIn$_5$. Therefore, one may expect the loss of the magnetic order in the Gd diluted lattice at the same concentration as Tb ($x_c\sim 0.68$). This is the expected result for a 3D magnetic system where the order must disappear at the percolation limit $x_c$=0.68\cite{MagnPhTrans,Kato,Vojta1}. The susceptibility points to a critical concentration of $x_c\sim 0.4$. This value would suggest a 2D percolation limit ($x_{c,2D}=0.39$ \cite{Yonezawa}) for the magnetic response, which is clearly the case for Ce$_x$La$_{1-x}$RhIn$_5$ where $T_N$ can be well defined from both $\chi(T)$ or $C/T$ datasets. For the present series, it seems more like an artefact of the large susceptibility associated to the connected magnetic regions.

Finally, dilution in Nd$_{1-x}$La$_x$RhIn$_5$ series~\cite{raimundo7} seems to share characteristics of  both Tb (with Ising-like couplings) and Gd (with Heisenberg like couplings).
It also shows an Ising like behaviour with sharp transitions. However, the 3D percolation limit is partially hidden by a divergent susceptibility, that does not allow to follow the magnetic transition beyond $x=0.4$.

\section{Conclusions}

In summary, the results of the magnetic properties of La-substituted Gd$_{1-x}$La$_x$RhIn$_5$ ($x\leq 0.50$) antiferromagnetic compounds are discussed. While GdRhIn$_5$ can be modeled by a $J=7/2$ Heisenberg model, when La substitutes Gd the spin wave behaviour is conserved at low temperatures but the sharp magnetic transition is destroyed. The substitutional disorder in these compounds is described by a distribution of critical temperatures as a function of $x$. It determines the deviation of the $T_N$ decrease with $x$ from a mean field behaviour. This weak disorder can be simulated if we consider the existence of regions with ordering temperature lower than the bulk order temperature. For larger concentrations, but still below the percolation limit, the critical temperatures distribution suffers a strong change that is clearly seen in both the low temperature specific heat and magnetic susceptibility data. Contrary to the expected behaviour, these results are not compatible with a classical spin Heisenberg model. The studied series represents a simple 4$f$ ($L=0$) AFM system for the study of substitutional disorder effects and short range order in anti-ferromagnets, which are relevant to understand AFM quantum phase transitions\cite{Light,Vojta2}. Furthermore, for the physics of the $R$RhIn$_5$ family, our results confirm the relevance of CEF effects on the behaviour of the magnetic properties when $R$=Tb,Nd. Further dilution studies on Gd$_{1-x}$Y$_x$RhIn$_5$ would be valuable to shed light into the role of the disorder discussed in this work since Y$^{3+}$ has closer atomic size to Gd$^{3+}$ than La$^{3+}$ has.

DJG thanks fruitful discussions with B. Alascio, J. Sereni, V. Correa and S. Bustingorry at CAB. This work was supported by FAPEMIG-MG (APQ-02256-12),  FAPESP-SP (06/50511-8 and 12/04870-7), CAPES (BEX-14248131), CNPq (2010-EXA020 and 304649/2013-9) and Agencia Nacional de Promoci\'on Cient\'ifica y Tecnol\'ogica (PICT 2012-1069).

\section{References}
\bibliography{Bibliog}

\begin{thebibliography}{10}
\expandafter\ifx\csname url\endcsname\relax
  \def\url#1{\texttt{#1}}\fi
\expandafter\ifx\csname urlprefix\endcsname\relax\def\urlprefix{URL }\fi
\expandafter\ifx\csname href\endcsname\relax
  \def\href#1#2{#2} \def\path#1{#1}\fi

\bibitem{Sarrao1}
J.~L. Sarrao, J.~D. Thompson, J. Phys. Soc. Jpn 76 (2007) 051013.

\bibitem{pagliuso7}
P.~G. Pagliuso, N.~O. Moreno, N.~J. Curro, J.~D. Thompson, M.~F. Hundley, J.~L.
  Sarrao, Z.~Fisk, A.~D. Christianson, A.~H. Lacerda, B.~E. Light, A.~L.
  Cornelius, Phys. Rev. B 66 (2002) 054433.

\bibitem{VictorCorrea}
V.~F. Correa, L.~Tung, S.~M. Hollen, P.~G. Pagliuso, N.~O. Moreno, J.~C.
  Lashley, J.~L. Sarrao, A.~H. Lacerda, Phys. Rev. B 69 (2004) 174424.

\bibitem{Light}
B.~E. Light, R.~S. Kumar, A.~L. Cornelius, P.~G. Pagliuso, J.~L. Sarrao, Phys.
  Rev. B 69 (2004) 024419.

\bibitem{Alver}
U.~Alver, R.~G. Goodrich, N.~Harrison, D.~W. Hall, E.~C. Palm, T.~P. Murphy,
  S.~W. Tozer, P.~G. Pagliuso, N.~O. Moreno, J.~L. Sarrao, Z.~Fisk, Phys. Rev.
  B 64 (2001) 180402(R).

\bibitem{christianson2}
A.~D. Christianson, E.~D. Bauer, P.~Pagliuso, N.~O. Moreno, M.~F. Hundley,
  J.~L. Sarrao, Physica B 312-313 (2002) 241.

\bibitem{wei3}
W.~Bao, A.~D. Christianson, P.~G. Pagliuso, J.~L. Sarrao, J.~D. Thompson, A.~H.
  Lacerda, J.~W. Lynn, Physica B 312-313 (2002) 120.

\bibitem{Tanatar}
M.~A. Tanatar, J.~Paglione, S.~Nakatsuji, D.~G. Hawthorn, E.~Boaknin, R.~W.
  Hill, F.~Ronning, M.~Sutherland, L.~Taillefer, C.~Petrovic, P.~C. Canfield,
  Z.~Fisk, Phys. Rev. Lett. 95 (2005) 067002.

\bibitem{Petrovic3}
C.~Petrovic, S.~L. Budko, V.~G. Kogan, P.~C. Canfield, Phys. Rev. B 66 (2002)
  054534.

\bibitem{Satoro}
S.~Nakatsuji, S.~Yeo, L.~Balicas, Z.~Fisk, P.~Schlottmann, P.~G. Pagliuso,
  N.~O. Moreno, J.~L. Sarrao, J.~D. Thompson, Phys. Rev. Lett. 89 (2002)
  106402.

\bibitem{Leticie}
L.~M. Ferreira, T.~Park, V.~Sidorov, M.~Nicklas, E.~M. Bittar, R.~Lora-Serrano,
  E.~N. Hering, S.~M. Ramos, M.~B. Fontes, E.~Baggio-Saitovich, H.~Lee, J.~L.
  Sarrao, J.~D. Thompson, P.~G. Pagliuso, Tuning the pressure-induced
  superconducting phase in doped cerhin5, Phys. Rev. Lett. 101 (2008) 017005.
\newblock \href {http://dx.doi.org/10.1103/PhysRevLett.101.017005}
  {\path{doi:10.1103/PhysRevLett.101.017005}}.

\bibitem{wei4}
W.~Bao, G.~Aeppli, J.~W. Lynn, P.~G. Pagliuso, J.~L. Sarrao, M.~F. Hundley,
  J.~D. Thompson, Z.~Fisk, Phys. Rev. B 65 (2002) 100505(R).

\bibitem{Takeuchi2006}
N.~V. Hieu, H.~Shishido, T.~Takeuchi, A.~Thamizhavel, H.~Nakashima,
  K.~Sugiyama, R.~Settai, T.~D. Matsuda, Y.~Haga, M.~Hagiwara, K.~Kindo,
  Y.~Onuki, J. Phys. Soc. Jpn 75 (2006) 074708.

\bibitem{raimundo5}
R.~Lora-Serrano, D.~J. Garcia, E.~Miranda, C.~Adriano, C.~Giles, J.~G.~S.
  Duque, P.~G. Pagliuso, Phys. Rev. B. 79 (2009) 024422.

\bibitem{raimundo7}
R.~Lora-Serrano, D.~J. Garcia, E.~Miranda, C.~Adriano, L.~Bufai\c{c}al,
  J.~G.~S. Duque, P.~G. Pagliuso, Physica B 404 (2009) 3059.

\bibitem{vojta3}
T.~Vojta, Rare region effects at classical, quantum and nonequilibrium phase
  transitions, J. Phys. A: Math. Gen. 39 (2006) R143--R205.

\bibitem{MagnPhTrans}
L.~J. de~Jongh, Static thermodynamic properties of site-random magnetic system:
  and the percolation problem, in: M.~Ausloss, R.~J. Elliott (Eds.), Magnetic
  Phase Transitions: Proceedings of a Summer School at the Ettore Majorana
  Centre, Springer-Verlag, 1983.

\bibitem{Kato}
K.~Kato, S.~Todo, K.~Harada, N.~Kawashima, S.~Miyashita, H.~Takayama, Phys.
  Rev. Lett. 84 (2000) 4204.

\bibitem{Vojta1}
T.~Vojta, J.~Schmalian, Phys. Rev. Lett. 95 (2005) 237206.

\bibitem{Yonezawa}
F.~Yonezawa, S.~Sakamoto, M.~Hori, Phys. Rev. B 40 (1989) 636.

\bibitem{pagliuso3}
P.~G. Pagliuso, J.~D. Thompson, M.~F. Hundley, J.~L. Sarrao, Z.~Fisk, Crystal
  structure and low-temperature magnetic properties of rmmin3m+2 compounds
  (m=rh or ir; m=1,2; r=sm or gd), Phys. Rev. B 63 (2001) 054426.
\newblock \href {http://dx.doi.org/10.1103/PhysRevB.63.054426}
  {\path{doi:10.1103/PhysRevB.63.054426}}.

\bibitem{granado2}
E.~Granado, B.~Uchoa, A.~Malachias, R.~Lora-Serrano, P.~G. Pagliuso,
  H.~Westfahl, Phys. Rev. B 74 (2006) 214428.

\bibitem{pagliuso5}
P.~G. Pagliuso, D.~J. Garcia, E.~Miranda, E.~Granado, R.~Lora-Serrano,
  C.~Giles, J.~G.~S. Duque, R.~R. Urbano, C.~Rettori, J.~D. Thompson, M.~F.
  Hundley, J.~L. Sarrao, J. Appl. Phys. 99 (2006) 08P703.

\bibitem{christianson1}
A.~D. Christianson, E.~D. Bauer, J.~M. Lawrence, P.~S. Riseborough, N.~O.
  Moreno, P.~G. Pagliuso, J.~L. Sarrao, J.~D. Thompson, E.~A. Goremychkin,
  F.~R. Trouw, M.~P. Hehlen, R.~J. McQueeney, Phys. Rev. B 70 (2004) 134505.

\bibitem{JensenAndMackintosh}
J.~Jensen, A.~R. Mackintosh, Rare Earth Magnetism, Oxford University Press, NY,
  1991.

\bibitem{Facio2015}
J.~I. Facio, D.~Betancourth, P.~Pedrazzini, V.~F. Correa, V.~Vildosola, D.~J.
  Garcia, P.~S. Cornaglia, Why the co-based 115 compounds are different: The
  case study of gdmin5 (m = co,rh,ir), Phys. Rev. B 91 (2015) 014409.
\newblock \href {http://dx.doi.org/10.1103/PhysRevB.91.014409}
  {\path{doi:10.1103/PhysRevB.91.014409}}.

\bibitem{Fisk1}
Z.~Fisk, J.~L. Sarrao, J.~L. Smith, J.~D. Thompson, Proc. Natl. Acad. Sci. USA
  92 (1995) 6663.

\bibitem{Holland}
T.~J.~B. Holland, S.~A.~T. Redfern, Mineralogical Magazine 61 (1997) 65.

\bibitem{Bacci}
S.~Bacci, E.~Gagliano, E.~Dagotto, Phys. Rev. B 44 (1991) 285.

\bibitem{Sushkov}
O.~P. Sushkov, J.~Oitmaa, Z.~Weihong, Phys. Rev. B 63 (2001) 104420.

\bibitem{DBetancourth1}
D.~G. D.~Betancourth, V.F.~Correa, Evidence of low energy anisotropy in
  gdcoin5, J. Low Temp. Phys. 179~(1-2) (2015) 90--93.
\newblock \href {http://dx.doi.org/10.1007/s10909-014-1233-2}
  {\path{doi:10.1007/s10909-014-1233-2}}.

\bibitem{ALPS}
A. W. Sandvik, \textit{Phys. Rev. B} \textbf{59}, 14157 (1999); F. Alet, S.
  Wessel, and M. Troyer \textit{Phys. Rev. E} \textbf{71}, 036706 (2005); L.
  Pollet, S. M. A. Rombouts, K. Van Houcke, and K. Heyde, \textit{Phys. Rev. E}
  \textbf{70}, 056705 (2004). M. Troyer et al., \textit{Lecture Notes in
  Computer Science}, Vol. 1505, p. 191 (1998). See also
  https://alps.comp-phys.org/.

\bibitem{BlundellMagnetism}
S.~Blundell, Magnetism in condensed matter (Oxford master series in condensed
  matter physics)., Oxford University Press, 2001.

\bibitem{Vojta2}
T.~Vojta, Quantum griffiths effects and smeared phase transitions in metals:
  Theory and experiment, J. Low Temp. Phys. 161 (2010) 299--323.

\end{thebibliography}

\end{document}